\documentclass[aps,prc,twocolumn,showpacs,floatfix,nofootinbib,a4paper]{revtex4}
\usepackage{graphicx} 
\usepackage{dcolumn} 
\usepackage{bm}         
\usepackage{color}

\newcommand{\be}{\begin{equation}}
\newcommand{\ee}{\end{equation}}
\newcommand{\ba}{\begin{eqnarray}}
\newcommand{\ea}{\end{eqnarray}}
\newcommand{\bd}{\begin{displaymath}}
\newcommand{\ed}{\end{displaymath}}

\begin{document}
\title{Fluctuations in Hadronizing QGP}

\author{
 L.P. Csernai$^{1,2}$, G. Mocanu$^3$ and  Z. N\'eda$^3$}

\affiliation{
$^1$
Department of Physics and Technology,
University of Bergen, 5007, Bergen, Norway\\
$^2$
Wigner Research Centre for Physics,
1525 Budapest, Pf. 49, Hungary\\
$^3$
Department of Physics, Babe\c{s}-Bolyai University, Cluj, Romania
}
\date{\today}

\begin{abstract}
The dynamical development of the cooling and 
hadronizing quark-gluon Plasma (QGP)
is studied in a simple model assuming critical fluctuations in the
QGP to Hadronic Matter (HM) and a first order transition in a small finite 
system. We consider an earlier determined free-energy density curve 
in the neighbourhood of the critical point, with two local minima
corresponding to the equilibrium hadronic and QGP configurations.
In this approach the divergence at $e = 0$ eliminates fluctuations
with negative or zero energy. The barrier between
the equilibrium states is 
obtained from an estimated value of the surface tension between 
the two phases.
We obtain a characteristic behavior for the skewness and the kurtosis
of energy density fluctuations, which can be studied via
a beam energy scan program.
\end{abstract}


\maketitle

\section{Introduction}
\label{intro}

In central heavy ion collisions fluctuations may occur due to critical
phenomena arising from a phase transition in the Equation of State (EoS).
Fluctuations arising from the initial asymmetry are smaller in these 
head-on reactions. In the neighborhood of the critical point the shear 
viscosity of 
the quark-gluon Plasma (QGP) is becoming small 
\cite{CKM}, which also facilitates the appearance 
of fluctuations.

The flow effects depend on the initial state profile and on the
viscosity. Turbulence appears only for small viscosity, which indicates
the critical point of the matter \cite{CKM}, and it is a sensitive
measure of viscosity and its minimum at the critical point.

Molecular dynamics and Fluid Dynamics simulations of heavy ion collisions
suggest that collective flow asymmetries can be measured \cite{Amelin91},
both if the global asymmetry or random flow arising from the initial 
state transverse momentum fluctuations or longitudinal
center of mass rapidity, $y_{CM}$, fluctuations are causing it.
These alternative sources may
also lead to specific statistical characteristics as discussed 
in ref. \cite{Wang12}. These phenomena contribute to a spatial spread
of the matter and energy density variations are present even if
we do not have a phase transition in our EoS. 

Here, on the other hand, we study random fluctuations of thermodynamical
origin caused by the phase transition based on the considerations described 
in ref. \cite{CsN94}.
The field is under intensive recent study, and several aspects
of the phase transitions have discussed the non-Gaussian fluctuations
\cite{intro}.

For a realistic reaction model we have to describe the final
stage of the reaction also. In order to study the properties
of the phase transition we need to have sufficient energy to
form QGP in a sufficiently large volume, and then the system
must hadronize. At relatively low energies this hadronization
might happen well before the freeze out (FO) stage, and then
we have only rather indirect information about the phase transition.

In our recent studies  \cite{Wang12} we use
a Multi Module Model or Hybrid Model approach to describe high energy
heavy ion collisions in the RHIC and LHC energy range.
Then from the locally equilibrated QGP we have to form hadrons.
We do not assume that the hadronization happens in chemical equilibrium
as this would take too long time \cite{CK92} and would not allow for
baryons of high strangeness.

In high energy heavy ion reactions around 50 GeV/nucleon or above
when QGP is formed, we have low mass quarks (5-10 MeV) and massless
gluons in large numbers and the conserved baryon charge has a minor
effect. On the hadronic side also mainly mesons and baryon-antibaryon 
pairs are formed. 
The observable phase transition happens
when the plasma expands, hadronizes, and freezes out. At this stage
the system may be in the vicinity of the critical point so critical
fluctuations may appear. In a finite volume this would show up as
energy density fluctuation, which would then lead to charged hadron
number fluctuation also (and much less in net baryon charge
fluctuations).

In the present simple model we assume two coexisting phases 
in a finite system near a first order
phase transition. We study the abundance of 
the energy density distribution of the two 
phases in the mixed phase domain in terms of the volume ratio.

This dynamical development is not directly observable, nevertheless,
we can observe the FO moment, which should be similar for 
different locations in the system.

At the same time we can vary the beam energy, which shifts the
FO point versus the critical point, so a beam energy scan can
provide us the series of information we are interested in.

\section{Critical Fluctuations}
\label{s2}

Following ref. \cite{CsN94} we briefly present the way
to describe critical fluctuations
following the ideas of the Landau-Ginsburg theory for critical phenomena.  

In case of QGP to hadronic matter (HM) transition
the essential difference is that in phase equilibrium the
energy density of the HM phase is much lower than that of
the QGP phase. As a consequence the usual 4th order
polynomial to describe the free energy of the system is not
realistic as it would lead to considerable population
of negative energy density states, which would be unphysical.

This problem was solved in ref. \cite{CsN94}, where the
4th order polynomial approach was substituted by a Laurent series.

Using the simple bag model EoS the equilibrium
value of the energy density in the low-temperature,
low-energy-density phase (HM) denoted
by $e_h$, and the equilibrium value of the energy-density
in the high-temperature, high-energy-density, QGP, phase
denoted by $e_q$ can be calculated as
\begin{equation}
e_h(T) = \frac{\pi^2}{10(\hbar c)^3} T^4
\label{eh}
\end{equation}
and
\begin{equation}
e_q(T) = \frac{\pi^2}{(\hbar c)^3}
\left( \frac{37}{30} T^4 + \frac{34}{90} T^4_c \right) .
\label{eq}
\end{equation}
To study the fluctuations we find the free energy,
 $f(e)$, for arbitrary values of $e$, and not
just for the energy densities $e_q$ and $e_h$. For
these equilibrium points:
\begin{equation}
f(e_q(T)) = -p_q(T) = - \frac{\pi^2}{90(\hbar c)^3} (37 T^4- 34 T^4_c),
\label{pq}
\end{equation}
\begin{equation}
f(e_h(T)) = -p_h(T) = - \frac{\pi^2}{30(\hbar c)^3} T^4.
\label{ph}
\end{equation}
Following the Landau theory we approximate now the free energy density as a
polynomial in the neighborhood of an $e_0$ energy density 
($e_0 \in [e_h,e_q]$), where it has a local maximum. In order to obtain the
required divergence at $e = 0$, a slightly modified functional form  is 
assumed:
\begin{equation}
f(e)  =f_1 + \frac{K_1}{e} + K_2 (e-e_0) +K_3(e-e_0)^2 + K_4(e-e_0)^3 .
\end{equation}
The constants, $f_1,\, K_1,\, K_2,\, K_3,\, K_4$ 
can be determined from thermodynamic considerations 
as done in ref. \cite{CsN94}. One will obtain

\begin{eqnarray}
K_1=\frac{\sigma}{\xi_0} \frac{1}{A_0 A_1},
\label{K1} \\
K_2=\frac{K_1}{e_0^2},\\
K_3=-\frac{K_1}{2}\frac{e_h^2(e_q+e_0)+e_q^2(e_0+e_h)-e_0^2(e_q+e_h)}{e_q^2e_h^2e_0^2},\\
K_4=\frac{K_3}{3} \frac{e_he_q+e_qe_0+e_0e_h}{e_q^2e_h^2e_0^2},
\end{eqnarray}
where
\begin{eqnarray}
A_0=\frac{(e_q-e_0)(e_0-e_h)}{e_0^3e_h^2e_q^2} \cdot(2e_qe_h+e_0e_q+e_he_0),\\
A_1=\frac{(e_q-e_h)^2}{3 \sqrt{2}}+\frac{[e_0-(e_q+e_h)/2]^2\sqrt{\pi}}{2}.
\end{eqnarray}
In Eq. (\ref{K1}) $\sigma$ represents the surface tension 
of a hadronic bubble and $\xi_0$ is the
characteristic size of a hadronic droplet. Once the $K_i$ values are 
expressed as a function of $e,e_0,e_q,e_h,\xi_0,\sigma$ values, 
the unknown $f_1$ and $e_0$ parameters 
are obtainable at any temperatures from Eqs (\ref{pq}) and (\ref{ph}).
Following our previous work \cite{CsN94}, we have used 
the $T_c=0.169MeV$, $\xi_0=3fm$ and $\sigma=0.05GeV/fm^2$ values in all
our calculations.

\begin{figure}[ht] 
\begin{center}
\includegraphics[width=3.4in]{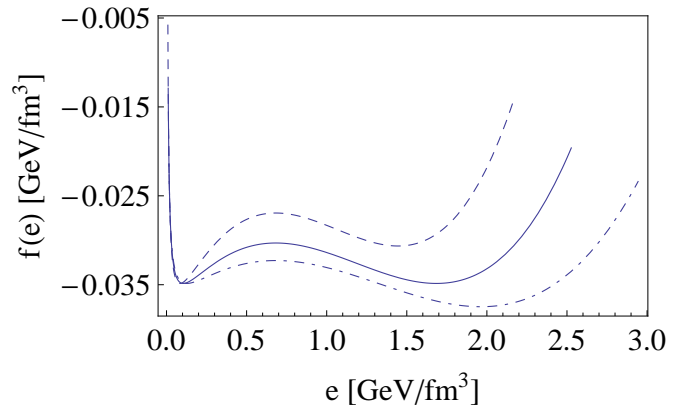}
\caption{
(color online)
The free energy density as a function of energy density $e$, for $T=0.95T_c$ 
(dashed), $T=T_c$ (full line) and $T=1.05 T_c$ (dot-dashed).
}
\label{Fig-1}
\end{center}
\end{figure}

For temperatures in the vicinity of the critical temperature the free energy 
density curve as a function of the energy density 
is illustrated in Fig. \ref{Fig-1}. 

\begin{figure}[ht] 
\begin{center}
\includegraphics[width=3.4in]{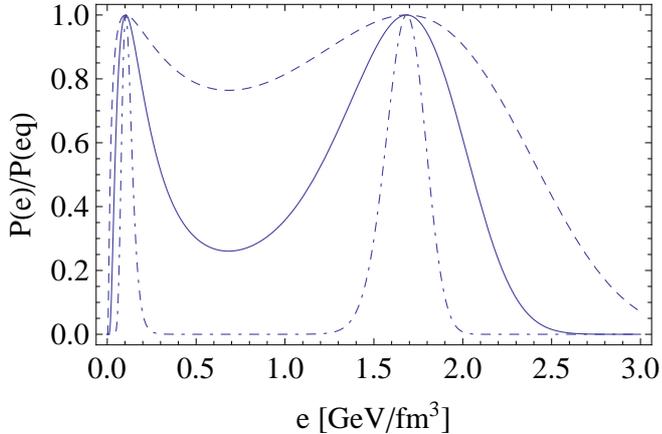}
\caption{
(color online)
The relative probability of finding a state of a given energy density $e$ 
for $T=T_c$, in a system of volume 
$\Omega = 10\mbox{ fm}^3$ (dashed),  
$\Omega = 50\mbox{ fm}^3$ (full line) and  
$\Omega = 500\mbox{ fm}^3$ (dot-dashed).
 }
\label{Fig-2}
\end{center}
\end{figure}

\begin{figure}[ht] 
\begin{center}
\includegraphics[width=3.4in]{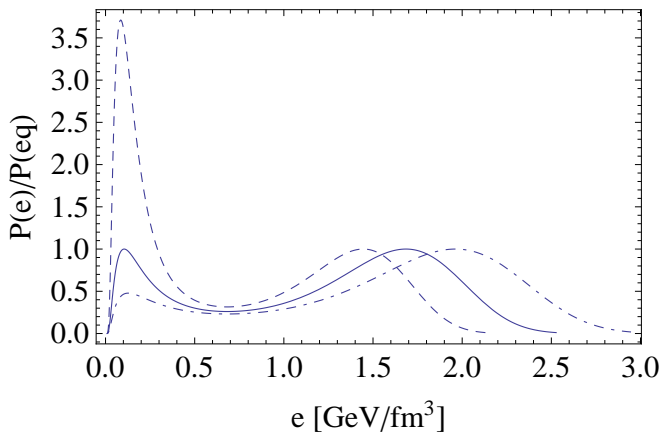}
\caption{
(color online)
The relative probability of finding a state of a given energy density $e$ 
for the $T=0.95 T_c$ (dashed), $T=T_c$ (continuous) and 
$T=1.05 T_c$ (dot-dashed) temperatures.
The of volume of QGP is $\Omega =  500\mbox{ fm}^3$.
 }
\label{Fig-3}
\end{center}
\end{figure}

Once the free energy curve is known, one can estimate the probability density
of finding the system in a state with energy density $e$: 
$P(e)\propto exp(-\beta F(e))$, where
$F(e)=\Omega f(e)$, with $\Omega$ the volume of the created QGP. On Fig.
\ref{Fig-2} we plot at the critical temperature the  
characteristic $P(e)/P(e_q)$ curves, considering different volumes for the 
QGP ($\Omega=10 \mbox{ fm}^3$, $50 \mbox{ fm}^3$, and $500 \mbox{ fm}^3$ 
values). Also, in Fig.
\ref{Fig-3} we show the $P(e)/P(e_q)$ curves for $\Omega=500 \mbox{ fm}^3$ and
the same temperatures as in Fig. \ref{Fig-1}.

It is also important to mention that different thermodynamical
parameters (especially intensives and extensive ones) do not have to
show the same critical fluctuation properties, so we have to
study the fluctuations of several parameters. Furthermore, the
statistical physics estimates assume a single thermal source
at or near the critical point, while we estimate here also the
effects of spatial fluctuations, which arise from a dynamically
expanding fluid flow even in the least fluctuating configuration.

%

\section{Hadronization and Expansion}

The dynamically developing flow pattern leads to a spatial distribution
of all thermodynamical quantities, while the system expands rapidly.
Finally the supercooled QGP can hadronize
rapidly and almost simultaneously it freezes out. This final
stage of the reaction can be described by a non-equilibrium model.

We assume central collisions only, to avoid the effects from azimuthal
flow asymmetries and from particle emission from projectile and target
residues (spectator evaporation) \cite{Amelin91}.

In a theoretical approach we can assume a spatial distribution of the
$x$ quantity.
For the variable $x$ the averages and various order moments
distributions can be written as
\begin{equation}
\langle x^{n} \rangle=\int{x^n P(x) dx},
\end{equation}
\begin{eqnarray}
M^{(n)} &=& \langle (x-\langle x \rangle)^n \rangle= \\
\nonumber
 &=& \int{(x-\langle x \rangle)^n P(x) dx},
\end{eqnarray}
where $P(x)$ is the spatial distribution weighted, e.g.,
by the baryon charge density in the center of mass frame (CF).
The spatial variance, the skewness and the kurtosis can be obtained
from these moments:
\begin{eqnarray}
\Delta x = \langle (x-\langle x \rangle)^2 \rangle = M^{(2)} \ ,
\label{variance}
\end{eqnarray}
\begin{eqnarray}
S = \frac{\langle (x-\langle x \rangle )^3 \rangle}{(\Delta x)^{3/2}} 
  = \frac{M^{(3)}}{(M^{(2)})^{3/2}} \ , 
\label{skewness}
\end{eqnarray}
\begin{eqnarray}
K = \frac  {\langle (x- \langle x \rangle )^4 \rangle }{(\Delta x)^2}-3 
  = \frac{ M^{(4)}}{(M^{(2)})^2}-3 \ .
\label{kurtosis}
\end{eqnarray}

By using these averages, first we can calculate specific extensives,
which are governed by strict conservation
laws. The total baryon charge, energy and momentum conservations
are governed by the continuity
equation and by the relativistic Euler equation.
\ba
N^\mu,_\mu &=& 0 , \\
T^{\mu\nu},_\nu &=& 0 ,
\ea
and as a consequence, the total momentum in the CF
should remain zero during the development,
while the average specific energy per net nucleon number
\be
\langle \varepsilon^{CF} \rangle \equiv \frac{T^{00}}{N^0} = const.
\ee
should remain constant in CF.

The total baryon number, $N_{tot}$,
is exactly conserved in the reaction. At the same time
the average baryon charge density is decreasing.

\section{Skewness and Kurtosis}

In this section we study the skewness and kurtosis of the specific energy
density and consequently the charged particle densities according to 
Eq.~(\ref{skewness}) and Eq.~(\ref{kurtosis}). We determine these 
quantities as a function of the
systems temperature and also as a function of the volume abundance ($r_h$) 
of the hadronic matter.
We assume that in a rapid transition, where critical fluctuations dominate
and the two phases are not separated these two phases are in thermal
equilibrium. The temperature decreases rapidly starting from QGP where
$T>T_c$ until the hadronization completes at $T < T_c$.
As the phases are not separated the simplest estimate
for any temperature $T$, is that the volume abundance of the hadronic 
matter is:
\begin{equation}
r_h=\frac{P(e_h)}{P(e_q)+P(e_h)}, 
\end{equation}
where $e_h$ and $e_q$ are the energy densities of the pure phases
defined in eqs. (\ref{eh}, \ref{eq}), and the probability densities,
$P(e)$, are defined in section \ref{s2}. This relation makes a one to one
correspondence between the volume abundance and the equilibrium temperature
in a rapidly expanding and cooling system during the process of a
phase transition.

\begin{figure}[ht] 
\begin{center}
\includegraphics[width=3.4in]{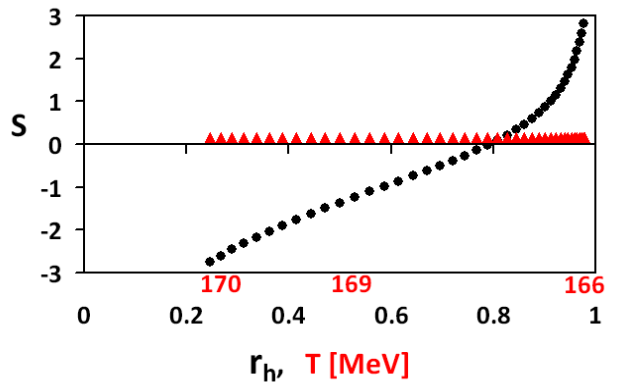}
\caption{
(color online)
Skewness as a function of the volume abundance of the hadronic matter 
(denoted as $r_h$, where $1$ represents complete hadronization). 
The temperature scale is 
also indicated for clarity, the identifiers represent increments 
of $0.1$ MeV in $T$. Results for $\Omega = 500 fm^3$.
}
\label{Fig-4}
\end{center}
\end{figure}

The skewness, Fig.  \ref{Fig-4}, is first negative
(indicating a longer tail on the lower energy side), 
then at 80 \% HM ($r_h=0.8$) volume abundance it turns into positive 
(indicating a longer tail on the high energy side). The hadronization
can be parametrized both as a function of the volume abundance of the
growing hadronic phase or the decreasing equilibrium temperature
of the system with critical fluctuations.

\begin{figure}[ht] 
\begin{center}
\includegraphics[width=3.4in]{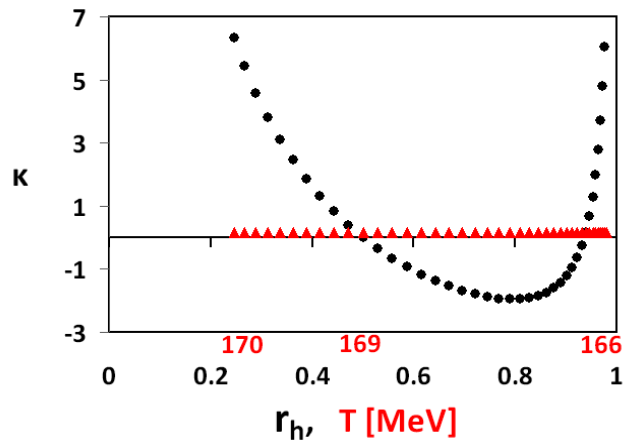}
\caption{
(color online)
Kurtosis as a function of the volume abundance of the hadronic matter 
(denoted as $r_h$, where $1$ represents complete hadronization). The 
temperature scale is also indicated for clarity, the identifiers 
represent increments of $0.1$ MeV in $T$. Results for $\Omega = 500 fm^3$.
 }
\label{Fig-5}
\end{center}
\end{figure}

In Fig. \ref{Fig-5}, we can see that the kurtosis is positive at first,
then turns to be negative (the distribution becomes wider) in the 
phase transition domain, while it becomes positive again
as the phase transition completes. The minimum of kurtosis is at 80 \%
HM volume abundance. Notice that the kurtosis is increasing much
sharper on the hadronic side, which is a clear consequence of the
energy difference between the two phases.  This asymmetry appears as 
a result of the Laurent series expansion, and it would not show up with
the usual 4th order polynomial approximation!

In a multi-module or hybrid-model construction
(e.g where the PIC hydro stage is attached to a parton and hadron
cascade model PACIAE), the flow features are matched \cite{Yun10}
to a subsequent dynamical model, which describes dynamical, non-equilibrium,
rapid hadronization. This type of models can describe realistically
the statistical properties in a dynamical phase transition, which
determine the hadron distribution in the final stage.  

This stage would then explicitly describe the random fluctuations
arising both from the phase transition and the flow dynamics.

In ref. \cite{Daimei-CPOD} a mixed particle method is
introduced, which could separate the fluctuations arising from local
critical fluctuations.
The "mixed events" are actually eliminating two particle
correlations, and only the single particle distributions
remain. Thus for central events these are mainly local
correlations, which may arise from local fluctuations caused
by energy and baryon charge clustering in a phase transition.
The method separates the consequences of such correlations.

This method can be used both in hybrid model calculations and
in experiments, to separate the fluctuation effects form the
collective flow and the phase transition dynamics.

\section{Conclusions}

In this work we reiterated earlier results on critical fluctuations
\cite{CsN94}
in the phase transition between quark-gluon plasma and hadronic matter.
This model is specific because of the large difference of the
energy density of the two phases, where $e_q \gg e_h$.
Here this work is extended to the evaluation of frequently used
statistical parameters like the Skewness and Kurtosis of typical parameters
like charged particle multiplicities.  In continuum models
the energy density fluctuations or baryon charge fluctuations
could carry the same role, however at ultra relativistic energies
a large number of antiparticles are created therefore the
energy density fluctuations are better representing the total 
charge particle multiplicity fluctuation.

Our model is furthermore extended to study the dynamical change of
these typical parameters during the hadronization process, where
the development is very characteristic and informative for the
phase transition we study.

In experiments one can measure these parameters (together with
all other measurable quantities) at the Freeze Out (FO) time 
(or at the FO hypersurface).  Luckily the recent RHIC Beam Energy Scan
program, scans the same statistical parameters at a series of 
different beam energies, which also provide a set of different FO
points mapping the dynamics of the phase transition or the most
interesting part of it.  The FO point at high energies where 
QGP is formed may be simultaneous to the phase transition, as
hadronization must always take place in these reactions. At lower
beam energies when QGP is not formed the FO is in the hadronic phase.
At very high energies where high energy density QGP is formed in a 
large volume, the system may have sufficiently large volume and 
large energy density that FO happens later then the hadronization,
i.e. in the pure hadronic state. 

The dynamical changes of the statistical parameters this way can
provide valuable information at what stage of the phase transition
we are at a given reaction.

{\bf Acknowledgment} 
Work supported from the research grant
PNII/PCCE 312/2008 on Complex Ideas. GM is supported by the 
project POSDRU/107/1.5/S/76841.



\begin{thebibliography}{99}

\bibitem{CKM}
  L.P.~Csernai, J.I.~Kapusta and L.D.~{McLerran},
  Phys.~Rev.~Lett.~{\bf 97}, 152303 (2006).

\bibitem{Amelin91}
  N.S. Amelin, E.F. Staubo, L.P. Csernai, V.D. Toneev, K.K. Gudima,
  Phys. Rev. C {\bf 44}, 1541 (1991).

\bibitem{Wang12}
  D.J. Wang et al.,
  Submitted to Eur. Phys. J A (2012) 

\bibitem{CsN94}
  L.P. Csernai and Z. Neda, Phys. Lett. B 337, 25 (1994).

\bibitem{intro}
  M. A. Stephanov, Phys. Rev. Lett. 102, 032301 (2009);
   R.V. Gavai and S. Gupta, Phys. Rev. D 78, 114503 (2008);
  Tapan K. Nayaka for the STAR collaboration,
  Nucl. Phys. A 830,  555c-558c   (2009);
  Y. Zhou, S.S. Shi, K. Xiao, K.J. Wu, and F. Liu,
  Phys. Rev. C 82, 014905 (2010);
  M Nahrgang, T Schuster, M Mitrovski, R Stock and M Bleicher,
  J. Phys. G 38, 124150 (2011).

\bibitem{CK92}
  L.P. Csernai and J.I. Kapusta, Phys. Rev. Lett. {\bf 69}, 737 (1992); and\\
  L.P. Csernai and J.I. Kapusta, Phys. Rev. D {\bf 46}, 1379 (1992).

\bibitem{CsMSS2011}
  L.P. Csernai, V.K. Magas, H. St{\"o}cker and D.D. Strottman,
  Phys. Rev. C {\bf 84},  024914 (2011).

\bibitem{CSA11}
  L.P. Csernai, D.D. Strottman and Cs. Anderlik,
  arXiv:1112.4287v1 [nucl-th].

\bibitem{MCs001}
  V.K.~Magas, L.P.~Csernai, and D.D. Strottman,
  Phys. Rev. C {\bf 64} (2001) 014901; Nucl. Phys. A {\bf 712}, 167 (2002).

\bibitem{Huovinen}
  P. Huovinen and P. Petreczky, J. Phys. G: {\bf 38}, 124103 (2011).

\bibitem{PACIAESa}
  Ben-Hao Sa, Dai-Mei Zhou, Yu-Liang Yan, Xiao-Mei Li,
  Sheng-Qin Feng, Bao-Guo Dong and Xu Cai, Comp. Phys. Commun., {\bf 183},
  333 (2012); arXiv: 1104.1238v1 [nucl-th].

\bibitem{Cleymans}
  J. Cleymans, H. Oeschler, K. Redlich  and S. Wheaton,
  Phys. Rev. C {\bf 73}, 034905 (2006).

\bibitem{Horvat}
  Sz. Horvat, V.K. Magas, D.D. Strottman, L.P. Csernai,
  Phys. Lett. B {\bf 692}, 277 (2010).

\bibitem{cser}
L. P. Csernai, Z. Neda, Phys. Lett. {\bf B 337}, 25 (1994).


\bibitem{Yun10}%
  Yun Cheng, L.P. Csernai, V. K. Magas, B.R. Schlei and D. Strottman, 
Phys. Rev. C {\bf  81}, 064910 (2010).

\bibitem{Daimei-CPOD}
  Dai-Mei Zhou, et al., presentation at the Int. Workshop on
  Critical Point and Onset of Deconfinement (CPOD), 7 - 11 November 2011,
  Wuhan, China. (http://conf.ccnu.edu.cn/~cpod2011/agenda.html)


\end{thebibliography}
\end{document}